 \documentclass[12pt,preprint]{aastex}

\usepackage{graphicx}
\usepackage{amsmath}

\def \apj {ApJ}

\def \apjl {ApJL}
\def \solphys {Solar Phys.}

\def \aap {A\&A}

\def \deg {$^{\rm o}$ }

\newcommand{\citeN}[1]{\citeauthor{#1} (\citeyear{#1})}
\newcommand{\citeNP}[1]{\citeauthor{#1} \citeyear{#1}}

\shortauthors{Socas-Navarro \& Elmore}
\shorttitle{Multi-line Observations of Spicules}

\begin{document}

\title{Physical Properties of Spicules from Simultaneous Spectro-Polarimetric
  Observations of He~I and Ca~II Lines}

\author{H. Socas-Navarro\thanks{Visiting Astronomer, National Solar
   	Observatory,  operated 
   	by the Association of Universities for Research in Astronomy, 
	Inc. (AURA), under cooperative agreement with the National Science
	Foundation.}
}
\author{D. Elmore$^{\rm 1}$}
   	\affil{High Altitude Observatory, NCAR\thanks{The National Center
	for Atmospheric Research (NCAR) is sponsored by the National Science
	Foundation.}, 3450 Mitchell Lane, Boulder, CO 80307-3000, USA}
	\email{navarro@ucar.edu}

\date{}%

\begin{abstract}
We present full Stokes observations from SPINOR in the \ion{Ca}{2} infrared
triplet and the \ion{He}{1} multiplet at 1083~nm from which some properties
of spicules have been derived. There are important advantages in multi-line
observations, particularly from different elements. We find that the
orientation of the plane of polarization is very different for the Ca and He
lines, which provides the 
first direct model-independent evidence for magnetic fields in
spicules. Our data shows that the Ca and He lines have almost identical
widths. Since the Ca atom is 10 times heavier than He, we are able to
conclude that 
most of the broadening is non-thermal ($\simeq$16~km~s$^{-1}$) and to set 
an upper limit of 13~kK to the spicular temperatures. The bisectors of the
lines span a velocity range of over 15~km~s$^{-1}$ for the He line and
30~km~s$^{-1}$ for the Ca ones. The vertical gradient of line-of-sight
velocities is also very different for both elements. We obtain
2.8~km~s$^{-1}$~Mm$^{-1}$ from He versus 6.4~km~s$^{-1}$~Mm$^{-1}$ from
Ca. These properties, and others from similar observations, should be taken
into account in future physical models of spicules.
\end{abstract}


\section{Introduction}
\label{sec:intro}

Spicules are elongated, nearly-vertical structures visible in emission
above the solar limb in chromospheric lines. First observed over 100 years
ago (\citeNP{S1877}), spicules have thus far eluded our efforts to understand
their physical nature and origin. Once viewed as a very attractive problem for
solar physicists, interest seemed to fade away somewhat since the early
1970s. Most of the knowledge gathered until then, mainly from spectroscopic
observations of H$\alpha$, He D$_3$ or Ca~H and~K, is compiled in a
comprehensive review by \citeN{B68}. Since then, very few works have been
published with new observational properties of spicules and some authors have
concentrated instead in novel modeling efforts (the most recent by
\citeNP{DPEJ04}).  

However, this scenario seems to be about to change drastically. Modern
spectro-polarimeters capable of observing chromospheric lines off of the
solar limb are providing new observational constrains that show considerable 
promise to expand our knowledge on spicules (and other chromospheric
structures). In addition to the present work, there are two other papers in
press with new spectro-polarimetric observations of \ion{He}{1} lines, namely
the 1083~nm multiplet (\citeNP{TBMC+04}) and D$_3$
(\citeNP{LAC04}). Hopefully, this surge is only the beginning of a
renewed interest in spicule observations. New data is needed both to inspire
and to constrain future theoretical models.

We present here for the first time observations of spicules in the
\ion{Ca}{2} infrared triplet (in particular the 849.8 and 854.2~nm lines),
alongside the \ion{He}{1} multiplet at 1083~nm. The dataset analyzed here was
recorded using the new SPINOR (Spectro-Polarimeter for INfrared and Optical
Regions) instrument, at the Dunn Solar Telescope (DST) in the Sacramento Peak
Observatory. We believe that multi-line 
observations from different elements (ideally with very different atomic
weights) have important advantages to determine some magnetic, thermal and
dynamic parameters of spicules. For example, we present evidence of the
existence of a magnetic field in spicules without having to resort on
complex Hanle effect modeling. In this sense, our work can be considered
the first model-independent verification of the magnetic nature of spicules
(although such fields have been obtained from Hanle effect
measurements by \citeNP{TBMC+04}; \citeNP{LAC04}). Since the thermal
broadening depends strongly on the atomic weight, multi-line observations are
also helpful to set boundaries to this type of broadening.

\section{Observations}

The observations that we report on were obtained on 19 June 2004 with
the following SPINOR configuration. We used a spectrograph grating with 308
lines/mm. The ASP cameras (256 by 256 pixels) were set to record 
the \ion{Ca}{2} lines at 849.8 and 854.2~nm. The new SPINOR Pluto camera (488
by 652 pixels) was set to the \ion{He}{1} multiplet at 1083~nm. 

The spectrograph slit had a width of 80~$\mu$m, which corresponds to an
angular size of 0.6'' on the DST focal plane. Short scans of 10 steps were
observed at various positions on the solar limb. The scanning step was chosen
to be 0.22'', which is significantly smaller than the typical angular
resolution ($\sim$1''). This was done so that it would be possible to bin
multiple images in the scanning direction, thus increasing the signal to
noise ratio without saturating any of the detectors. This approach is useful
for multi-line observations, particularly when significant variations exist
among the quantum efficiencies of the cameras or the photon flux they
receive. In this case, the ASP cameras are observing a spectral region
where they have a quantum efficiency of $\sim$30\%, whereas the Pluto camera
has only $\sim$3\% efficiency near 1~$\mu$m.

The polarization modulator used in this run was not the new SPINOR achromatic
bycristaline modulator, which we had not yet received from the vendor at the
time of these observations\footnote{
The new polarization modulator is now available to users of the DST upon
request. 
}.
We were able to use the ASP modulator which, although not optimal, still
provides some modulation at these wavelengths. The linear polarization
efficiency obtained with our setup was $\simeq$26\% at the wavelength of the
Ca lines and $\simeq$15\% at the wavelength of the He multiplet. For circular
polarization we obtained 36\% and 17\%, respectively.

The main limitation of our observations is probably the low signal level
in the 1083~nm region, due to the combination of low detector and
polarimeter efficiency discussed above. Nevertheless, there is still
sufficient signal especially in the bright emission cores of the lines
to allow for a detailed study of some important spicule properties.

An occulting disk at the telescope prime focus was inserted to block the
bright photospheric emission. This is helpful to reduce the amount of stray
light from the solar disk which would otherwise contaminate the relatively
faint spicule emission. The downside of this technique is that it
complicates the data reduction and calibration by removing some references,
such as quiet Sun intensity or telluric lines (useful for absolute
wavelength determination). The polarization calibration also becomes more
difficult because we no longer have continuum radiation to use as a source of
unpolarized light.

The telescope was calibrated as explained in \citeN{SNEL04},
using an array of achromatic polarizers on top of the entrance
window. Calibration data were obtained simultaneously for the entire
wavelength spectrum by means of a cross-dispersor in the beam before the
detector. The datasets thus obtained were fitted with a wavelength-dependent
model of the DST, allowing us to determine the optical
parameters that characterize the telescope polarimetric response. Another set
of achromatic polarization optics at the DST exit port was used to determine
the Mueller matrix of the instrument. 

Even after the exhaustive calibration outlined above, 
some residual cross-talk still remains from Stokes~$I$ into $Q$, $U$ and 
$V$. In practice, this residual is usually determined from the
continuum polarization (preferably at disk center, especially when observing
at blue wavelengths where scattering polarization may be important). This was
not possible in our observations because we only had spicule line emission
(the photospheric continuum was blocked by the occulting 
disk). Fortunately, 
we also had disk observations of active regions from the same day. The
spicule data reported here were taken only minutes away from an earlier
sunspot observation. We used the residual cross-talk measured in the disk
data to correct the spicule dataset.

\begin{figure*}
\plotone{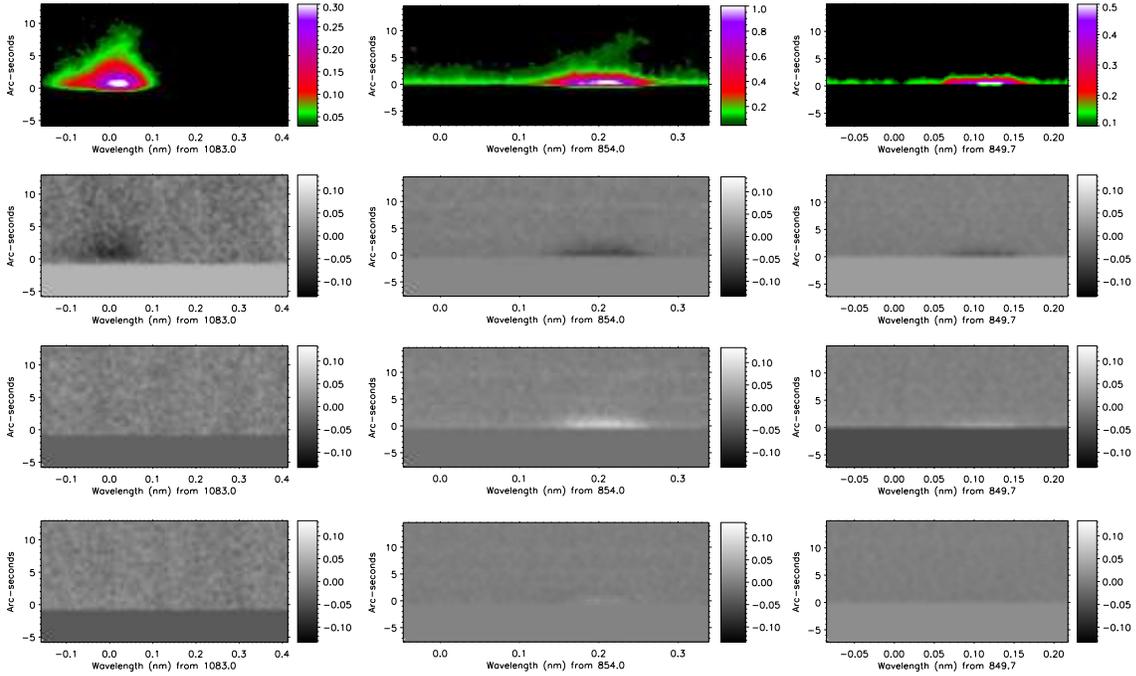}
\caption{
Stokes spectra of a spicule observed in the \ion{He}{1} multiplet (left), and
the \ion{Ca}{2} lines at 854.2~nm (middle) and 849.8 (right). The Stokes
parameters displayed, from top to bottom, are $I$, $Q$, $U$ and $V$,
respectively. The 
vertical scale is arc-seconds measured from the edge of the occulting
disk. All profiles are normalized to the maximum intensity of the 854.2~nm
line.
\label{fig:profs}
}
\end{figure*}

Fig~\ref{fig:profs} shows the (polarized) spectra of a spicule at the three
spectral regions, observed on the East limb of the Sun at -70\deg
latitude. Notice the absence of signal in the Stokes~$V$ 
images (bottom row), which indicates that no residual cross-talk remains
above the noise level. The real Stokes~$V$ signal one would expect to have is
too weak to show in these images (\citeNP{TBMC+04}). 

\section{Results}

Our observations show a bright emission core concentrated in the lower part
of the image. Above this bright emission there is a diffuse faint tail that
extends upwards and exhibits a strong redshift that increases
with height. The brightest emission is seen in the core of 
854.2. Interestingly, while 1083 has the weakest emission near the limb,
the bright core extends higher than either one of the Ca lines (up to
$\simeq$5'' in He, as opposed to $\simeq$2'' in Ca). The emission
in 849.8 is weaker than in 854.2 and fades away very rapidly with height. The
faint redshifted tail is not detectable in this line with our noise level.

Most of the discussion in this paper refers to the bright area near the limb
(the lower $\simeq$2'' in 849.8 and 854.2, and $\simeq$5'' in 1083), as
this is the region where we have stronger signals. 

\begin{figure*}
\plottwo{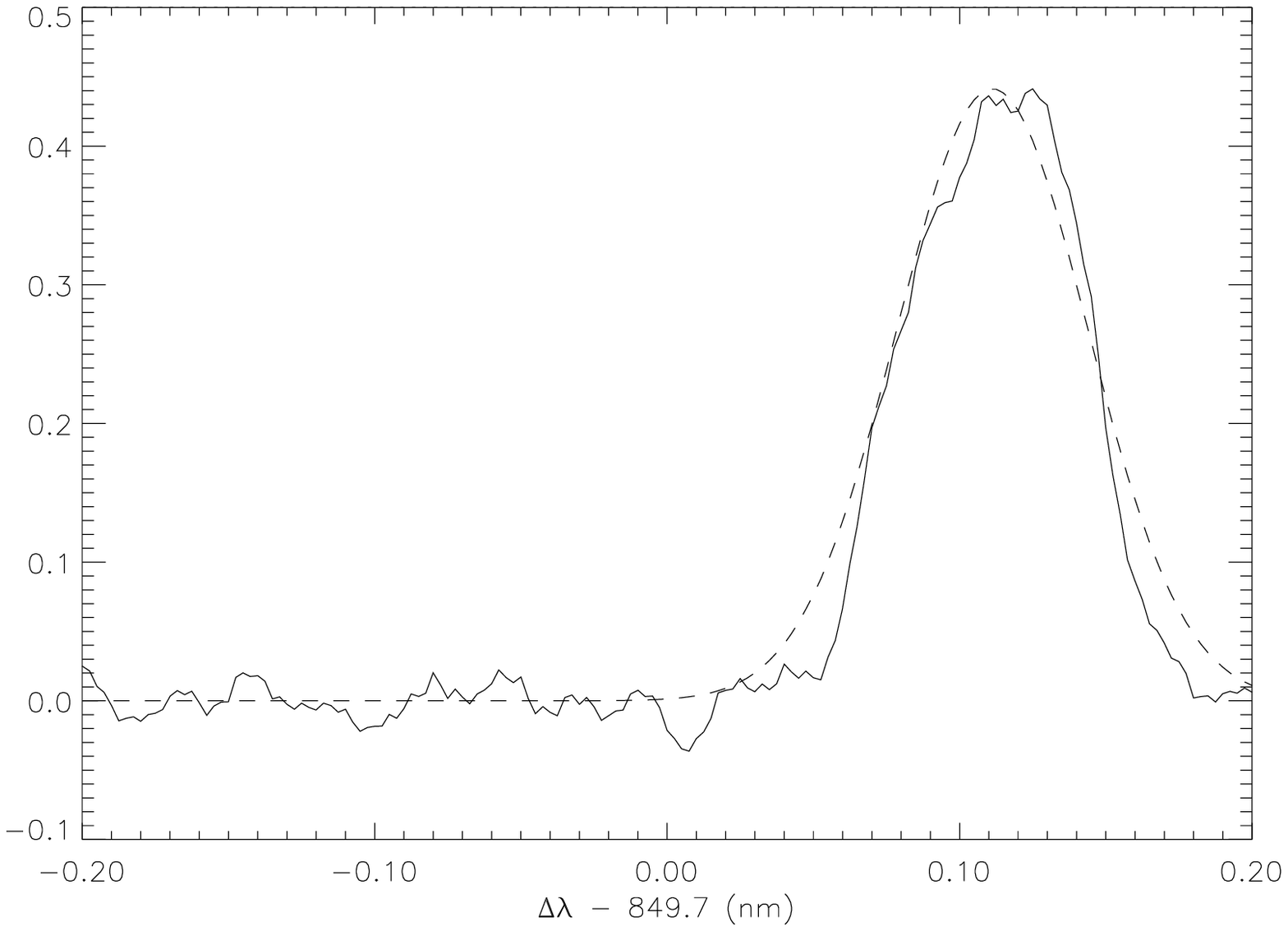}{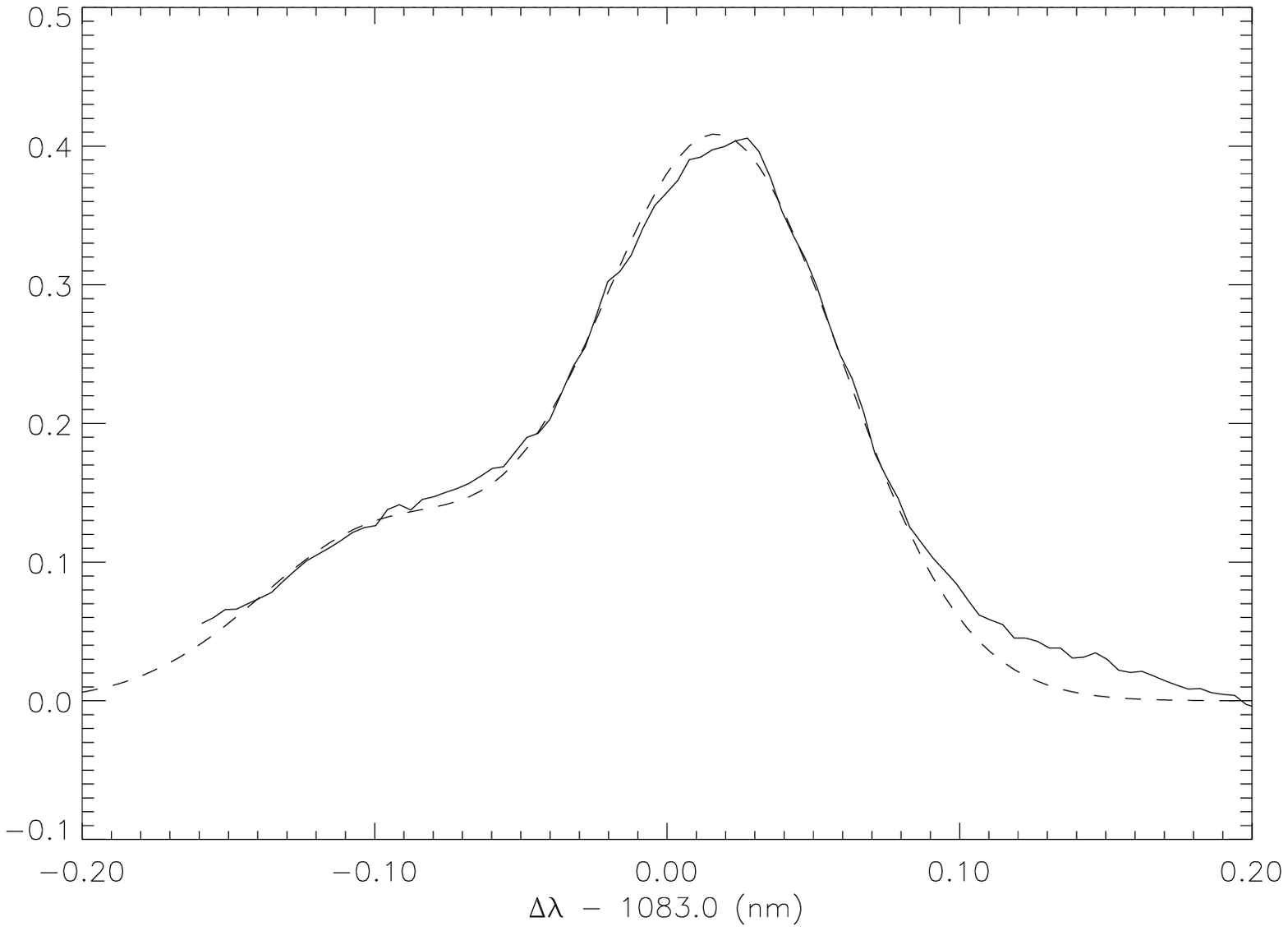}
\caption{Gaussian fits to the observed \ion{Ca}{2} line at 849.8~nm (left)
  and the 
  \ion{He}{1} multiplet at 1083~nm (right). The \ion{He}{1} profile is
  fitted with two Gaussians of the same width centered at 1082.91 and
  1083.03~nm, respectively. The fitted widths are: 18 km~s$^{-1}$ for 854.2
  (not shown), and 16.6 km~s$^{-1}$ for both 849.8 and 1083. Both observed
  and synthetic profiles are normalized to the maximum 
  intensity of the 854.2~nm line. The fits overestimate the
  wings of the Ca lines, but underestimate those of the He lines. Notice the
  relative Doppler shifts from the wings to the
  core of the Ca line. 
\label{fig:fits34}
}
\end{figure*}

Let us start considering the Doppler widths of the
lines. Fig~\ref{fig:fits34} shows the result of fitting Gaussian curves to
two of the spectral lines observed. The Ca atom is 10 times heavier than
He. If the lines are broadened by microscopic thermal velocities, one would
expect the He line to be over 3 times broader than those of Ca. However, this
is 
not the case as all three lines have very similar widths. This indicates that
the thermal broadening accounts only for a small part of the total
line width. This conclusion is still valid even if the lines have some amount
of optical thickness (which we shall neglect in this work), since this has a
relatively little impact on the Doppler width (\citeNP{TBMC+04}),
insufficient to explain the factor 3 discrepancy. 

If we assume that the non-thermal broadening mechanism (rotation, expansion,
etc) has the same effect on the three lines at a given spatial 
position, it is possible to establish an upper limit to the electron
temperature of the spicule. Let us separate the total width as:
\begin{equation}
\Delta \lambda^2=\Delta \lambda_T^2 + \Delta \lambda_N^2 \, ,
\end{equation}
where $\Delta \lambda_T$ and $\Delta \lambda_N$ denote the
thermal and non-thermal broadening, respectively. As we increase the
temperature, $\Delta \lambda_T$ increases more rapidly for the He lines than
for the Ca ones:
\begin{equation}
\Delta \lambda_T^{He}=\sqrt{10} \Delta \lambda_T^{Ca} \, ,
\end{equation}
which implies that the $\Delta \lambda_N$ also decreases more rapidly for
He. If we 
assume a typical uncertainty of 1.5~km~s$^{-1}$ in $\Delta \lambda$
(taken from the difference between the fits to the two Ca lines), we can
conclude that $T < 13$~kK. For higher temperatures, $\Delta
\lambda_N$ would be considerably smaller in the He lines than in those of Ca,
which contradicts our initial assumption that $\Delta \lambda_N^{Ca}=\Delta
\lambda_N^{He}$. A summary of these calculations is 
shown in Table~\ref{table}. 

\begin{deluxetable}{cccc}
  \tablewidth{0pt}
  \tablecaption{
  \label{table}}
  \tablehead{   & \ion{He}{1} 1083~nm & \ion{Ca}{2} 854.2~nm  & \ion{Ca}{2} 849.8~nm \\
}
  \startdata
$\Delta \lambda$ (km~s$^{-1}$) &   16.6 & 18.0 & 16.6  \\
$\Delta \lambda_N$ (km~s$^{-1}$), $T=10$~kK  & 15.3 & 17.9 & 16.5 \\
$\Delta \lambda_N$ (km~s$^{-1}$), $T=13$~kK  &  14.9 & 17.9 & 16.5 \\
$\Delta \lambda_N$ (km~s$^{-1}$), $T=15$~kK  &  14.6 & 17.8 & 16.4 \\
$< {d v / d h} >$ (km~s$^{-1}$~Mm$^{-1}$) & 2.8  &  6.4  & N.A.  \\
$U/Q$ ratio   &  0.07  & -1.29 &  -0.88 \\
 \enddata
\end{deluxetable}

Unfortunately, it is not possible to use a similar argument to obtain a
lower limit. The reason is that, once $\Delta \lambda_T$ becomes small
enough, the total broadening is dominated by the non-thermal component. In
this regime, all three lines end up having the same width and are compatible
with the observations regardless of the temperature. 

It is also interesting to consider the far wings of the
lines. \citeN{TBMC+04} noticed that their He observations had very extended
wings when compared to those of a Gaussian. Our observations confirm their
results for the He line (see Fig~\ref{fig:fits34}, right panel). However, we
find that the Ca lines show the opposite behavior. Both Ca lines have wings
that are actually {\it narrower} than those of a Gaussian (849.8 is shown in
Fig~\ref{fig:fits34}, left). These results indicate that the far wings are
probably dominated by the non-thermal broadening. It is not clear to us why
the Ca and He wings would exhibit such a different behavior.

\begin{figure*}
\plotone{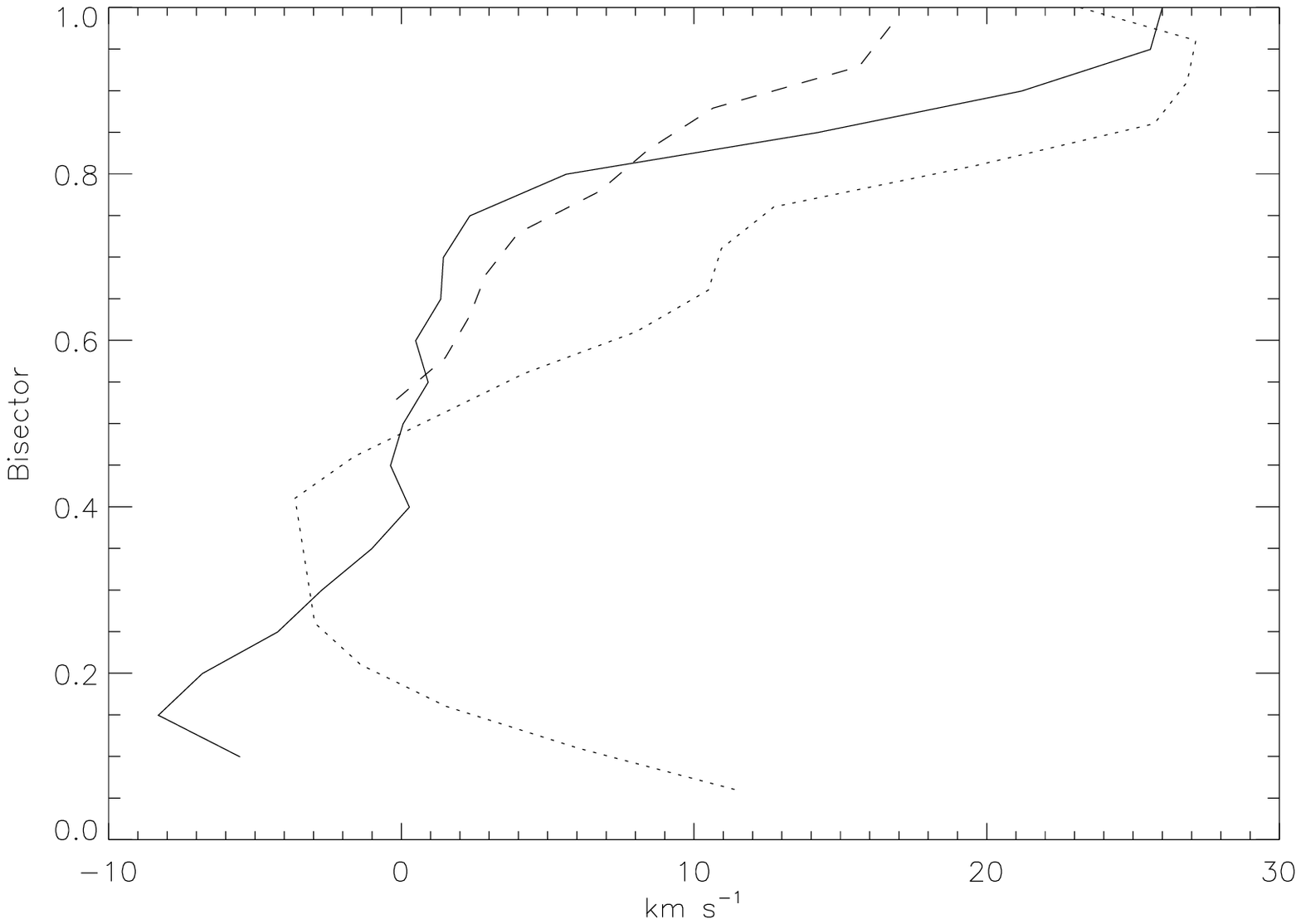}
\caption{Bisectors of the lines at 854.2 (solid), 849.8 (dotted) and 1083
  (dashed). Positive abscissa values correspond to redshifts. The zero point
  for all three lines is arbitrary due to the lack 
  of a suitable absolute wavelength reference in the data.
\label{fig:bisec}
}
\end{figure*}

The observed profiles shown in Fig~\ref{fig:fits34} (as well as that of the
854.2~nm line, not shown in the figure) exhibit signatures of differential
Doppler shifts from the wings to the core of the lines. To study this in some
more detail we calculated the bisector of all three lines (shown in
Fig~\ref{fig:bisec}). For the 1083~nm line we have left out the lower part of
the curve, to avoid the contamination introduced by the blend with the blue
component of the multiplet. All three lines exhibit a strong redshift from
approximately half of the profile to the core. This redshift is stronger in
the Ca lines, where the relative velocities reach 30~km~s$^{-1}$. The He
line, on the other hand, reaches only some 17~km~s$^{-1}$. The lower part of
the 849.8 curve is also redshifted, producing a C-shaped bisector. This is
not obvious in the 854.2 line, however. The interpretation of bisectors is
not as straightforward as it is in disk observations, in which the vertical
axis is directly related to height in the atmosphere. Here, a line profile
may be viewed more as a histogram of velocities. The intensity at a given
wavelength is related to the density of excited atoms with the corresponding
Doppler shift. These curves might be helpful as a test for future physical
models of spicules, in particular for the distribution of macroscopic
velocities within the emitting plasma.

Fig~\ref{fig:profs} also shows that the faint tail visible at 854.2 and
1083~nm above the bright emission exhibit a height-dependent redshift. We
have estimated the mean velocity gradient ($<d v / d h>$) associated with the
redshift in the range of heights between 5'' and 
7.5'' (approximately between 3700 and 5500~km above the limb). The results
are shown in Table~\ref{table} (fifth row). It is interesting to note that
the mean gradient differs by more than a factor 2 between the Ca and He
lines (recall that we are looking at the same spatial location along the
slit). Only 
\citeN{KK61} have reported a similar disparity between spicule 
velocities as measured in different lines (in their case, between H$\alpha$
and \ion{He}{1} D$_3$). This ``interesting phenomenon'', as
\citeN{B68} referred to it, would be very difficult to explain if one assumes
that both 
emissions come from the same region. Instead, we propose that the
line-of-sight integration is mixing together different structures in the
observed pixel. The excitation conditions in these structures may be
different, giving rise to stronger emission in either Ca or He and resulting
in different apparent velocities for each line. 

Let us now turn to the question of the possible presence of magnetic fields
in spicules. It is not straightforward to measure such fields ($\sim$10-30~G,
see \citeNP{TBMC+04}; \citeNP{LAC04}) because the 
Zeeman splitting is too small to be detected in the broad intensity profiles
of these lines, and the Stokes~$V$ signal is below the detection threshold
(or very close to it). Even if such Zeeman signals could be detected, the
lines would be formed in the so-called weak-field regime, in which it is not
possible to separate the intrinsic field strength from its filling factor in
the observed pixel. 

The only possible diagnostics left for reliable magnetic field
determinations is the Hanle effect. Still, the interpretation of Hanle
observations is not trivial. One needs a precise knowledge of the zero-field
polarization produced by the scattering of anisotropic radiation in the
higher atmosphere. The Hanle effect then leads (typically) to some
depolarization of the zero-field signal, as well as a rotation of the
linear polarization 
plane. In some sense, one essentially compares the theoretical zero-field
polarization with the observations and ascribes the difference to the effect
of a depolarizing magnetic field. It is therefore crucial to use a realistic
model of the 
atmosphere and the illumination, especially in cases like the 1083~nm lines
which are sensitive to coronal radiation.

To our best knowledge, the first claim that
spicules are magnetized is the recent work by \citeN{TBMC+04} (but see also
\citeNP{LAC04}), who carried 
out a detailed Hanle modeling of the 1083~nm multiplet. Having multiple
spectral lines in our observations allows us to confirm the presence of a
magnetic field in spicules using a straightforward, model-independent
reasoning. In the absence of a magnetic field, one would observe some degree
of scattering polarization. The plane of polarization of this signal is given
by the geometry of the scattering process and would be the same for all three
lines observed here. If, on the other hand, the scattering atoms are embedded
in a magnetic field, then the Hanle effect would rotate the plane of
polarization by an amount that depends on some parameters of the
transition. In this case, the orientation of the linear polarization would be
different in different lines. It is clear from Fig~\ref{fig:profs} that
the ratio of the Stokes $Q$ and $U$ signals is different for the Ca and He
lines, 
implying a different orientation of the polarization plane (and therefore the
presence of a magnetic field). The last row of
Table~\ref{table} lists the values of $U$/$Q$ averaged over the line core.

\section{Conclusions}

The observations that we present in this work provide some new insights into
the physical origin of spicules. By observing lines of both Ca and He we are
able to set constrains on the non-thermal broadening of the lines and the
temperature of the emitting plasma. The change in the orientation of the
polarization plane in the Ca and He lines is a convincing model-independent
evidence for magnetic fields in prominences. A quantitative determination of
the field strength and orientation would require detailed modeling of the
line formation, which is beyond the scope of the present work. The different
behavior of the emission wings in He and Ca (broader and narrower than a
Gaussian, respectively) is certainly intriguing. It seems appealing to
consider this as a clue for the nature of the non-thermal broadening
mechanism.

In spite of recent efforts, spicules still remain 
poorly understood and we feel that more data are urgently needed. We plan to
expand the results presented here in a forthcoming paper with new
observations and more detailed analyses. In particular we intend to use an
infrared camera and the new SPINOR modulator to obtain better signal-to-noise
ratios in the infrared (which, as stated above is one of our main limitations
here). Detailed Non-LTE modeling would be desirable and 
it seems to be well understood now, at least for the Ca lines both in the
Zeeman (\citeNP{SNTBRC00a}) and Hanle (\citeNP{MSTB01}) regimes. 

\acknowledgments

The authors are grateful to the Sac Peak observatory staff for their help
with the observations, in particular D. Gilliam, M. Bradford and
J. Elrod. Thanks are also due to R. Manso Sainz for suggesting a plausible
explanation for the difference between Ca and He velocity gradients, and also
for comments on an earlier draft of the manuscript.


\begin{thebibliography}{9}
\expandafter\ifx\csname natexlab\endcsname\relax\def\natexlab#1{#1}\fi

\bibitem[{{Beckers}(1968)}]{B68}
{Beckers}, J.~M. 1968, \solphys, 3, 367

\bibitem[{{De Pontieu} {et~al.}(2004){De Pontieu}, {Erd{\' e}lyi}, \&
  {James}}]{DPEJ04}
{De Pontieu}, B., {Erd{\' e}lyi}, R., \& {James}, S.~P. 2004, \nat, 430, 536

\bibitem[{{Krat} \& {Krat}(1961)}]{KK61}
{Krat}, V.~A., \& {Krat}, T.~V. 1961, Izvestiya Glavnoj Astronomicheskoj
  Observatorii v Pulkove, 22, 6

\bibitem[{{L\' opez Ariste} \& {Casini}(2004)}]{LAC04}
{L\' opez Ariste}, A., \& {Casini}, R. 2004, \aap, {\it submitted}

\bibitem[{{Manso Sainz} \& {Trujillo Bueno}(2001)}]{MSTB01}
{Manso Sainz}, R., \& {Trujillo Bueno}, J. 2001, in ASP Conf. Ser. 236:
  Advanced Solar Polarimetry -- Theory, Observation, and Instrumentation, 213

\bibitem[{{Secchi}(1887)}]{S1877}
{Secchi}, A. 1887, "Le Soleil, Vol. 2" (Gauthier-Villars, Paris)

\bibitem[{{Socas-Navarro} {et~al.}(2004){Socas-Navarro}, {Elmore}, \&
  {Lites}}]{SNEL04}
{Socas-Navarro}, H., {Elmore}, D., \& {Lites}, B. 2004, Solar Physics, {\it
  submitted}

\bibitem[{{Socas-Navarro} {et~al.}(2000){Socas-Navarro}, {Trujillo Bueno}, \&
  {Ruiz Cobo}}]{SNTBRC00a}
{Socas-Navarro}, H., {Trujillo Bueno}, J., \& {Ruiz Cobo}, B. 2000, \apj, 530,
  977

\bibitem[{{Trujillo Bueno} {et~al.}(2004){Trujillo Bueno}, {Merenda},
  {Centeno}, {Collados}, \& {Landi Degl'Innocenti}}]{TBMC+04}
{Trujillo Bueno}, J., {Merenda}, L., {Centeno}, R., {Collados}, M., \& {Landi
  Degl'Innocenti}, E. 2004, \apjl, {\it submitted}

\end{thebibliography}

\end{document}